\documentclass{article}
\usepackage{graphicx}
\usepackage{amsmath}

\input{tcilatex}

\begin{document}

\title{\textbf{Dual Gravitation }}
\author{W. Chagas-Filho \\
Physics Department, Federal University of Sergipe, Brazil}
\maketitle

\begin{abstract}
We propose a canonical relation between gravity and space-time
noncommutativity.
\end{abstract}

\section{Introduction}

\noindent\ The subject of space-time noncommutativity is now being the focus
of considerable interest. Space-time noncommutativity was first clearly
noticed in effective field theories obtained from string theory [1]. In this
context it involves a kind of constant noncommutativity which violates
Lorentz covariance. This is an unattractive feature and caused the interest
in this type of space-time noncommutativity to be rather limited.

A different kind of noncommutativity for the space-time coordinates appeared
in connection with generalizations of special relativity [2,3]. These
generalizations are based on principles of relativity which include, in
addition to the constant speed of light $c$, some other universal invariant,
like a minimum length or a constant radius of curvature. The space-time
noncommutativity associated to these generalizations of special relativity
is contained in generalizations of the Poincar\'{e} algebra, but it is not
clear if these generalized space-time algebras have a physical meaning.

The first Lorentz-covariant noncommutative space-time was proposed by H. S.
Snyder [4] in 1947. Because the Snyder commutators are based on a projective
geometry approach in momentum space to de Sitter space, this motivated the
desire to understand the relation between gravitation and space-time
noncommutativity. The search for this relation was further motivated by the
theoretical verification that, when quantum measurement processes involve
energies of the order of the Planck scale, the fundamental assumption of
locality is no longer a good approximation in theories containing gravity
[5]. The measurements alter the space-time metric in a fundamental manner
governed by the flat space-time commutation relation 
\begin{equation}
\lbrack x_{\mu },p_{\nu }]=i\eta _{\mu \nu }  \tag{1.1}
\end{equation}
and the classical field equations of gravitation [5]. This change in the
space-time metric destroys the locality, and hence the commutativity, of
position measurements operators [5]. Despite knowledge of these results, a
clear relation between gravitation and space-time noncommutativity was
lacking untill now. Researchers in the field usually take the working
hypothesis [6] that there is one physical property which at large scales
manifests itself as gravity, and at small scales as noncommutativity.

In this work we present formal manipulations which indicate that there is a
canonical relation between gravitation and space-time noncommutativity.
According to this canonical relation, in a noncommutative space-time, usual
gravitational fields, which depend on the space-time positions only, can not
exist. In a noncommutative space-time, only momentum-dependent gravitational
fields can exist. This new and surprising picture leads us to the concept of
dual gravitation. As we show here, the usual picture of a position-dependent
gravitational field defined over a commutative space-time can be obtained
from this new picture by performing a canonical duality transformation. The
results of this work therefore elucidate at least one of the many possible
relations between gravity and noncommutativity.

The paper is divided as follows. In section two we review the basics of
massless relativistic particle theory and show how the classical analogue of
Snyder's noncommutative quantized space-time can be constructed in particle
theory. We then discuss how the classical canonical brackets we found lead
to the concept of dual gravitation. In section three we extend massless
particle theory to a more symmetric theory in a higher dimensional
space-time and show how two gauge-equivalent sets of canonical brackets can
be constructed for this extended theory. These two dual sets of canonical
brackets give a clear picture of the relation between gravitation and
space-time noncommutativity in the extended theory and confirm the idea of
dual gravitation. Some concluding remarks appear in section four.

\section{Massless Relativistic Particles}

A massless relativistic particle in a $d$-dimensional Minkowski space-time
with signature $(d-1,1)$, where $d$ is the number of space-like dimensions,
is described by the action 
\begin{equation}
S=\frac{1}{2}\int d\tau \lambda ^{-1}\dot{x}^{2}  \tag{2.1}
\end{equation}
where a dot denotes derivatives with respect to the parameter $\tau $.
Action (2.1) is invariant under the local infinitesimal reparametrizations 
\begin{equation}
\delta x_{\mu }=\epsilon (\tau )\dot{x}_{\mu }  \tag{2.2a}
\end{equation}
\begin{equation}
\delta \lambda =\frac{d}{d\tau }[\epsilon (\tau )\lambda ]  \tag{2.2b}
\end{equation}
and therefore describes gravity on the world-line. Action (2.1) is also
invariant under the global Poincar\'{e} transformations 
\begin{equation}
\delta x^{\mu }=a^{\mu }+\omega _{\nu }^{\mu }x^{\nu }  \tag{2.3a}
\end{equation}
\begin{equation}
\delta \lambda =0  \tag{2.3b}
\end{equation}
where $\omega _{\nu \mu }=-\omega _{\mu \nu },$ under the global scale
transformations 
\begin{equation}
\delta x^{\mu }=\alpha x^{\mu }  \tag{2.4a}
\end{equation}
\begin{equation}
\delta \lambda =2\alpha \lambda  \tag{2.4b}
\end{equation}
where $\alpha $ is a constant, and under the conformal transformations 
\begin{equation}
\delta x^{\mu }=(2x^{\mu }x^{\nu }-\eta ^{\mu \nu }x^{2})b_{\nu }  \tag{2.5a}
\end{equation}
\begin{equation}
\delta \lambda =4\lambda x.b  \tag{2.5b}
\end{equation}
where $b_{\mu }$ is a constant vector. As a consequence of the presence of
these global invariances we can define in space-time the following field 
\begin{equation}
V=a^{\mu }p_{\mu }-\frac{1}{2}\omega ^{\mu \nu }M_{\mu \nu }+\alpha D+b^{\mu
}K_{\mu }  \tag{2.6}
\end{equation}
with the generators 
\begin{equation}
p_{\mu }  \tag{2.7a}
\end{equation}
\begin{equation}
M_{\mu \nu }=x_{\mu }p_{\nu }-x_{\nu }p_{\mu }  \tag{2.7b}
\end{equation}
\begin{equation}
D=x.p  \tag{2.7c}
\end{equation}
\begin{equation}
K_{\mu }=2x_{\mu }x.p-x^{2}p_{\mu }  \tag{2.7d}
\end{equation}
$p_{\mu }$ generates translations in space-time,$\ M_{\mu \nu }$ generates
space-time rotations, $D$ is the generator of space-time dilatations and $%
K_{\mu }$ generates conformal transformations. These generators define the
algebra 
\begin{equation}
\{p_{\mu },p_{\nu }\}=0  \tag{2.8a}
\end{equation}
\begin{equation}
\{p_{\mu },M_{\nu \lambda }\}=\eta _{\mu \nu }p_{\lambda }-\eta _{\mu
\lambda }p_{\nu }  \tag{2.8b}
\end{equation}
\begin{equation}
\{M_{\mu \nu },M_{\lambda \rho }\}=\eta _{\nu \lambda }M_{\mu \rho }+\eta
_{\mu \rho }M_{\nu \lambda }-\eta _{\nu \rho }M_{\mu \lambda }-\eta _{\mu
\lambda }M_{\nu \rho }  \tag{2.8c}
\end{equation}
\begin{equation}
\{D,D\}=0  \tag{2.8d}
\end{equation}
\begin{equation}
\{D,p_{\mu }\}=p_{\mu }  \tag{2.8e}
\end{equation}
\begin{equation}
\{D,M_{\mu \nu }\}=0  \tag{2.8f}
\end{equation}
\begin{equation}
\{D,K_{\mu }\}=-K_{\mu }  \tag{2.8g}
\end{equation}
\begin{equation}
\{p_{\mu }.K_{\nu }\}=-2\eta _{\mu \nu }D+2M_{\mu \nu }  \tag{2.8h}
\end{equation}
\begin{equation}
\{M_{\mu \nu },K_{\lambda }\}=\eta _{\nu \lambda }K_{\mu }-\eta _{\lambda
\mu }K_{\nu }  \tag{2.8i}
\end{equation}
\begin{equation}
\{K_{\mu },K_{\nu }\}=0  \tag{2.8j}
\end{equation}
computed in terms of the Poisson brackets 
\begin{equation}
\{p_{\mu },p_{\nu }\}=0  \tag{2.9a}
\end{equation}
\begin{equation}
\{x_{\mu },p_{\nu }\}=\eta _{\mu \nu }  \tag{2.9b}
\end{equation}
\begin{equation}
\{x_{\mu },x_{\nu }\}=0  \tag{2.9c}
\end{equation}
The algebra (2.8) is the conformal space-time algebra. The massless particle
theory defined by action (2.1) is a conformal theory in $d$ dimensions.

As is well known, conformal invariance in $d$ dimensions is equivalent to
Lorentz invariance in $d+2$ dimensions. By defining [7] 
\begin{equation}
L_{\mu \nu }=M_{\mu \nu }  \tag{2.10a}
\end{equation}
\begin{equation}
L_{\mu d}=\frac{1}{2}(p_{\mu }+K_{\mu })  \tag{2.10b}
\end{equation}
\begin{equation}
L_{\mu (d+1)}=\frac{1}{2}(p_{\mu }-K_{\mu })  \tag{2.10c}
\end{equation}
\begin{equation}
L_{d(d+1)}=D  \tag{2.10d}
\end{equation}
the conformal algebra (2.8) can be put in the standard form 
\begin{equation}
\{L_{MN},L_{RS}\}=\delta _{MR}L_{NS}+\delta _{NS}L_{MR}-\delta
_{MS}L_{NR}-\delta _{NR}L_{MS}  \tag{2.11}
\end{equation}
with $M,N=0,1,...,d,d+1$ and $\eta _{MN}=diag(-1,+1,...,+1,-1)$. This shows
that there are hidden dimensions in massless particle theory. In the next
section we will use these hidden dimensions to generalize the world-line
gravity action (2.1) to a more symmetric theory in a $(d+2)$-dimensional
space-time.

In the transition to the Hamiltonian formalism action (2.1) gives the
canonical momenta 
\begin{equation}
p_{\lambda }=0  \tag{2.12}
\end{equation}
\begin{equation}
p_{\mu }=\frac{\dot{x}_{\mu }}{\lambda }  \tag{2.13}
\end{equation}
and the canonical Hamiltonian 
\begin{equation}
H=\frac{1}{2}\lambda p^{2}  \tag{2.14}
\end{equation}
Equation (2.12) is a primary constraint [8]. Introducing the Lagrange
multiplier $\xi (\tau )$ for this constraint we can write the Dirac
Hamiltonian 
\begin{equation}
H_{D}=\frac{1}{2}\lambda p^{2}+\xi p_{\lambda }  \tag{2.15}
\end{equation}
Requiring the dynamical stability of constraint (2.12), $\dot{p}_{\lambda
}=\{p_{\lambda },H_{D}\}=0$, we obtain the secondary constraint 
\begin{equation}
\phi =\frac{1}{2}p^{2}\approx 0  \tag{2.16}
\end{equation}
Constraint (2.16) has a vanishing Poisson bracket with constraint (2.12),
being therefore a first-class constraint [8]. Constraint (2.12) generates
translations in the arbitrary variable $\lambda (\tau )$ and can be dropped
from the formalism.

In equation (2.16) we have introduced the \textbf{weak equality symbol }$%
\approx $. This is to emphasize that constraint $\phi $ is numerically
restricted to be zero in the subspace of phase space where the canonical
coordinates $(x^{\mu },p^{\mu })$ satisfy equation (2.16), but it does not
identically vanish throughout phase space. In particular, it has nonzero
Poisson brackets with the canonical positions. More generally, two functions 
$F$ and $G$ that coincide on the submanifold of phase space defined by the
constraint $\phi \approx 0$ are said to be \textbf{weakly equal }and one
writes $F\approx G$. On the other hand, an equation that holds throughout
phase space and not just on the submanifold $\phi \approx 0$ is called 
\textbf{strong}, and the usual equality symbol is used in that case. It can
be demonstrated that, in general [9] 
\begin{equation}
F\approx G\Leftrightarrow F-G=c_{i}(x,p)\phi _{i}  \tag{2.17}
\end{equation}

Now we point out that the massless particle Hamiltonian (2.14) is invariant
under the local scale transformations 
\begin{equation}
p_{\mu }\rightarrow \tilde{p}_{\mu }=\exp \{-\beta \}p_{\mu }  \tag{2.18a}
\end{equation}
\begin{equation}
\lambda \rightarrow \exp \{2\beta \}\lambda  \tag{2.18b}
\end{equation}
where $\beta $ is an arbitrary function of $x$ and $p$. From the equation
(2.13) for the canonical momentum we find that $x^{\mu }$ transforms as 
\begin{equation}
x^{\mu }\rightarrow \tilde{x}^{\mu }=\exp \{\beta \}x^{\mu }  \tag{2.18c}
\end{equation}
when $p_{\mu }$ transforms as in (2.18a). The local scale invariance (2.18)
of the massless particle Hamiltonian (2.14) is the residue of a broken gauge
invariance of action (2.1). The Lagrangian is not invariant because the
kinetic term $\dot{x}.p$ in the Legendre transformation, $L=\dot{x}.p-H$, is
not invariant under transformation (2.18). Perhaps the notion of broken
local scale invariance may be the clue for the quantum mechanics of the
gravitational field.

Consider now the bracket structure that transformations (2.18a) and (2.18c)
induce in the massless particle phase space. Retaining only the linear terms
in $\beta $ in the exponentials, we find that the new transformed canonical
variables $(\tilde{x}_{\mu },\tilde{p}_{\mu })$ obey the brackets 
\begin{equation}
\{\tilde{p}_{\mu },\tilde{p}_{\nu }\}=(\beta -1)[\{p_{\mu },\beta \}p_{\nu
}+\{\beta ,p_{\nu }\}]+\{\beta ,\beta \}p_{\mu }p_{\nu }  \tag{2.19a}
\end{equation}
\begin{equation*}
\{\tilde{x}_{\mu },\tilde{p}_{\nu }\}=(1+\beta )[\delta _{\mu \nu }(1-\beta
)-\{x_{\mu },\beta \}p_{\nu }]
\end{equation*}
\begin{equation}
+(1-\beta )x_{\mu }\{\beta ,p_{\nu }\}-\{\beta ,\beta \}x_{\mu }p_{\nu } 
\tag{2.19b}
\end{equation}
\begin{equation}
\{\tilde{x}_{\mu },\tilde{x}_{\nu }\}=(1+\beta )[x_{\mu }\{\beta ,x_{\nu
}\}-x_{\nu }\{\beta ,x_{\mu }\}]+\{\beta ,\beta \}x_{\mu }x_{\nu } 
\tag{2.19c}
\end{equation}
If we choose $\beta =\phi =\frac{1}{2}p^{2}\approx 0$ in equations (2.19)
and compute the brackets on the right side in terms of the Poisson brackets
(2.9), we find the expressions 
\begin{equation}
\{\tilde{p}_{\mu },\tilde{p}_{\nu }\}=0  \tag{2.20a}
\end{equation}
\begin{equation}
\{\tilde{x}_{\mu },\tilde{p}_{\nu }\}=(1+\frac{1}{2}p^{2})[\eta _{\mu \nu
}(1-\frac{1}{2}p^{2})-p_{\mu }p_{\nu }]  \tag{2.20b}
\end{equation}
\begin{equation}
\{\tilde{x}_{\mu },\tilde{x}_{\nu }\}=-(1+\frac{1}{2}p^{2})(x_{\mu }p_{\nu
}-x_{\nu }p_{\mu })  \tag{2.20c}
\end{equation}
We see from the above equations that, on the constraint surface defined by
equation (2.16), the brackets (2.20) reduce to 
\begin{equation}
\{\tilde{p}_{\mu },\tilde{p}_{\nu }\}=0  \tag{2.21a}
\end{equation}
\begin{equation}
\{\tilde{x}_{\mu },\tilde{p}_{\nu }\}=\eta _{\mu \nu }-p_{\mu }p_{\nu } 
\tag{2.21b}
\end{equation}
\begin{equation}
\{\tilde{x}_{\mu },\tilde{x}_{\nu }\}=-(x_{\mu }p_{\nu }-x_{\nu }p_{\mu }) 
\tag{2.21c}
\end{equation}
To impose $\phi =\frac{1}{2}p^{2}\approx 0$ strongly at the end of the
computation of brackets (2.20), the expression for the corresponding Dirac
brackets [8] on the right side should be used in place of the Poisson
brackets. However, for the special case $\beta =\phi =\frac{1}{2}%
p^{2}\approx 0$ we can use the property [9] of the Dirac bracket that, on
the first-class constraint surface, 
\begin{equation}
\{G,F\}_{D}\approx \{G,F\}  \tag{2.22}
\end{equation}
when $G$ is a first-class constraint and $F$ is an arbitrary function of the
canonical variables. This justifies the use of Poisson brackets to arrive at
(2.21).

Now, keeping the same order of approximation used to arrive at brackets
(2.19), that is, retaining only the linear terms in $\beta $, the
transformation equations (2.18a) and (2.18c) read 
\begin{equation}
\tilde{p}_{\mu }=\exp \{-\beta \}p_{\mu }=(1-\beta )p_{\mu }  \tag{2.23a}
\end{equation}
\begin{equation}
\tilde{x}_{\mu }=\exp \{\beta \}x_{\mu }=(1+\beta )x_{\mu }  \tag{2.23b}
\end{equation}
Using again the same function $\beta =\phi =\frac{1}{2}p^{2}\approx 0$ in
equations (2.23), we write them as 
\begin{equation}
\tilde{p}_{\mu }=p_{\mu }-\frac{1}{2}p^{2}p_{\mu }  \tag{2.24a}
\end{equation}
\begin{equation}
\tilde{x}_{\mu }=x_{\mu }+\frac{1}{2}p^{2}x_{\mu }  \tag{2.24b}
\end{equation}
or, equivalently, 
\begin{equation}
\tilde{p}_{\mu }-p_{\mu }=c_{\mu }(x,p)\phi  \tag{2.25a}
\end{equation}
\begin{equation}
\tilde{x}_{\mu }-x_{\mu }=d_{\mu }(x,p)\phi  \tag{2.25b}
\end{equation}
where $c_{\mu }(x,p)=-p_{\mu }$ and $d_{\mu }(x,p)=x_{\mu }$. Equations
(2.25) are in the form (2.17) and so we can write 
\begin{equation}
\tilde{p}_{\mu }\approx p_{\mu }  \tag{2.26a}
\end{equation}
\begin{equation}
\tilde{x}_{\mu }\approx x_{\mu }  \tag{2.26b}
\end{equation}
Using these weak equalities in brackets (2.21) we rewrite them as 
\begin{equation}
\{p_{\mu },p_{\nu }\}\approx 0  \tag{2.27a}
\end{equation}
\begin{equation}
\{x_{\mu },p_{\nu }\}\approx \eta _{\mu \nu }-p_{\mu }p_{\nu }  \tag{2.27b}
\end{equation}
\begin{equation}
\{x_{\mu },x_{\nu }\}\approx -(x_{\mu }p_{\nu }-x_{\nu }p_{\mu }) 
\tag{2.27c}
\end{equation}
to emphasize that these brackets are valid only on the constraint surface
defined by equation (2.16). In the transition to the quantum theory the
brackets (2.27) will reproduce the structure of the Snyder commutators.

Although the space-time coordinates now have non-vanishing classical
brackets, which will correspond to non-vanishing commutators in the
quantized theory, we can not say that an effective gravitational field
appears on the right side of bracket (2.27b), as would be expected from the
results in [5]. This is because, according to the current point of view, a
physical gravitational field should depend only on the particle's position.
We propose here that this point of view should be enlarged to contain also
the notion of momentum-dependent gravitational fields. This is because
bracket (2.27c) will unavoidably lead to space-time quantization, and a
position-dependent gravitational field could therefore never be a continuous
field. If we take the point of view that the resulting space-time geometry
can be determined from the gravitational contributions to the flat
commutator (1.1), we have to admit the possibility that in a noncommutative
space-time the gravitational field can only depend on the particle's
momentum, as is suggested by bracket (2.27b). Momentum which, according to
bracket (2.27a), remains continuous, giving therefore a continuous
momentum-dependent gravitational field. In the next section we will confirm
this interpretation by constructing the dual picture, that is, in a
commutative space-time the gravitational field must be position-dependent
because momenta are quantized.

As an initial step for the developments in the next section, we rewrite
(2.1) in the form 
\begin{equation}
S=\int d\tau (\dot{x}.p-\frac{1}{2}\lambda p^{2})  \tag{2.28}
\end{equation}
If we solve the equation of motion for $p_{\mu }$ that follows from action
(2.28) and insert the solution back into it, we recover action (2.1).

\section{Two-time physics}

The higher-dimensional extension of the massless particle action (2.28) is a
gauge theory with two time-like dimensions, usually refered to as ``
two-time physics'' [10-16]. The construction of this theory is based on the
introduction of a new gauge invariance in phase space, by gaugeing the
duality of the canonical commutator (1.1). This procedure leads to a
symplectic Sp(2,R) gauge theory. To remove the distinction between position
and momentum we rename them $X_{1}^{M}=X^{M}$ and $X_{2}^{M}=P^{M}$ and
define the doublet $X_{i}^{M}=(X_{1}^{M},X_{2}^{M})$. The local $Sp(2,R)$
symmetry acts as 
\begin{equation}
\delta X_{i}^{M}(\tau )=\epsilon _{ik}\omega ^{kl}(\tau )X_{l}^{M}(\tau ) 
\tag{3.1}
\end{equation}
$\omega ^{ij}(\tau )$ is a symmetric matrix containing three local
parameters and $\epsilon _{ij}$ is the Levi-Civita symbol that serves to
raise or lower indices. The $Sp(2,R)$ gauge field $A^{ij}$ is symmetric in $%
(i,j)$ and transforms as 
\begin{equation}
\delta A^{ij}=\partial _{\tau }\omega ^{ij}+\omega ^{ik}\epsilon
_{kl}A^{lj}+\omega ^{jk}\epsilon _{kl}A^{il}  \tag{3.2}
\end{equation}
The covariant derivative is 
\begin{equation}
D_{\tau }X_{i}^{M}=\partial _{\tau }X_{i}^{M}-\epsilon _{ik}A^{kl}X_{l}^{M} 
\tag{3.3}
\end{equation}
An action invariant under the $Sp(2,R)$ gauge symmetry is 
\begin{equation}
S=\frac{1}{2}\int d\tau (D_{\tau }X_{i}^{M})\epsilon ^{ij}X_{j}^{N}\eta _{MN}
\tag{3.4a}
\end{equation}
After an integration by parts this action can be written as 
\begin{equation*}
S=\int d\tau (\partial _{\tau }X_{1}^{M}X_{2}^{N}-\frac{1}{2}%
A^{ij}X_{i}^{M}X_{j}^{N})\eta _{MN}
\end{equation*}
\begin{equation}
=\int d\tau \lbrack \dot{X}.P-(\frac{1}{2}\lambda _{1}P^{2}+\lambda _{2}X.P+%
\frac{1}{2}\lambda _{3}X^{2})]  \tag{3.4b}
\end{equation}
where $A^{11}=\lambda _{3}$, $A^{12}=A^{21}=\lambda _{2}$, \ $A^{22}=\lambda
_{1}$ and the canonical Hamiltonian is 
\begin{equation}
H=\frac{1}{2}\lambda _{1}P^{2}+\lambda _{2}X.P+\frac{1}{2}\lambda _{3}X^{2} 
\tag{3.5}
\end{equation}
The equations of motion for the $\lambda $'s give the primary constraints 
\begin{equation}
\phi _{1}=\frac{1}{2}P^{2}\approx 0  \tag{3.6}
\end{equation}
\begin{equation}
\phi _{2}=X.P\approx 0  \tag{3.7}
\end{equation}
\begin{equation}
\phi _{3}=\frac{1}{2}X^{2}\approx 0  \tag{3.8}
\end{equation}
Constraints (3.6)-(3.8), as well as evidences of two-time physics, were
independently obtained in [7].

If we consider the Minkowski metric as the background space-time, we find
that the surface defined by the constraint equations (3.6)-(3.8) is trivial.
The only metric giving a non-trivial surface, preserving the unitarity of
the theory, and avoiding the ghost problem is the flat metric with two
time-like dimensions [10-16]. Following [10-16] we introduce another
space-like dimension and another time-like dimension and work in a Minkowski
space-time with signature $(d,2).$ Action (3.4b) is the $(d+2)$-dimensional
generalization of the $d$-dimensional massless particle action (2.28).
Action (3.4b) describes conformal gravity on the world-line.

We use the Poisson brackets 
\begin{equation}
\{P_{M},P_{N}\}=0  \tag{3.9a}
\end{equation}
\begin{equation}
\{X_{M},P_{N}\}=\eta _{MN}  \tag{3.9b}
\end{equation}
\begin{equation}
\{X_{M},X_{N}\}=0  \tag{3.9c}
\end{equation}
where $M,N=0,...,d+1$, and verify that constraints (3.6)-(3.8) obey the
algebra 
\begin{equation}
\{\phi _{1},\phi _{2}\}=-2\phi _{1}  \tag{3.10a}
\end{equation}
\begin{equation}
\{\phi _{1},\phi _{3}\}=-\phi _{2}  \tag{3.10b}
\end{equation}
\begin{equation}
\{\phi _{2},\phi _{3}\}=-2\phi _{3}  \tag{3.10c}
\end{equation}
These equations show that all constraints $\phi $ are first-class. Equations
(3.10) represent the symplectic $Sp(2,R)$ gauge algebra of two-time physics.

Action (3.4) also has a global symmetry under Lorentz transformations $%
SO(d,2)$ with generator [10-16] 
\begin{equation}
L^{MN}=\epsilon ^{ij}X_{i}^{M}X_{j}^{N}=X^{M}P^{N}-X^{N}P^{M}  \tag{3.11}
\end{equation}
It satisfies the space-time algebra (2.11) and is gauge invariant because it
has vanishing brackets with the first-class constraints (3.6)-(3.8), $%
\{L_{MN},\phi _{i}\}=0.$

Now, Hamiltonian (3.5) is invariant under the local scale transformations 
\begin{equation}
X^{M}\rightarrow \tilde{X}^{M}=\exp \{\beta \}X^{M}  \tag{3.12a}
\end{equation}
\begin{equation}
P_{M}\rightarrow \tilde{P}_{M}=\exp \{-\beta \}P_{M}  \tag{3.12b}
\end{equation}
\begin{equation}
\lambda _{1}\rightarrow \exp \{2\beta \}\lambda _{1}  \tag{3.12c}
\end{equation}
\begin{equation}
\lambda _{2}\rightarrow \lambda _{2}  \tag{3.12d}
\end{equation}
\begin{equation}
\lambda _{3}\rightarrow \exp \{-2\beta \}\lambda _{3}  \tag{3.12e}
\end{equation}
where $\beta $ is an arbitrary function of $X^{M}(\tau )$ and $P_{M}(\tau )$%
. Keeping only the linear terms in $\beta $ in transformation (3.12), we can
write the brackets 
\begin{equation}
\{\tilde{P}_{M},\tilde{P}_{N}\}=(\beta -1)[\{P_{M},\beta \}P_{N}+\{\beta
,P_{N}\}P_{M}]+\{\beta ,\beta \}P_{M}P_{N}  \tag{3.13a}
\end{equation}
\begin{equation*}
\{\tilde{X}_{M},\tilde{P}_{N}\}=(1+\beta )[\eta _{MN}(1-\beta
)-\{X_{M},\beta \}P_{N}]
\end{equation*}
\begin{equation}
+(1-\beta )X_{M}\{\beta ,P_{N}\}-X_{M}P_{N}\{\beta ,\beta \}  \tag{3.13b}
\end{equation}
\begin{equation}
\{\tilde{X}_{M},\tilde{X}_{N}\}=(1+\beta )[X_{M}\{\beta
,X_{N}\}-X_{N}\{\beta ,X_{M}\}]+X_{M}X_{N}\{\beta ,\beta \}  \tag{3.13c}
\end{equation}
for the transformed canonical variables. If we choose $\beta =\phi _{1}=%
\frac{1}{2}P^{2}$ $\approx 0$ in equations (3.13) and compute the brackets
on the right side using the Poisson brackets (3.10), we find the expressions 
\begin{equation}
\{\tilde{P}_{M},\tilde{P}_{N}\}=0  \tag{3.14a}
\end{equation}
\begin{equation}
\{\tilde{X}_{M},\tilde{P}_{N}\}=(1+\frac{1}{2}P^{2})[\eta _{MN}(1-\frac{1}{2}%
P^{2})-P_{M}P_{N}]  \tag{3.14b}
\end{equation}
\begin{equation}
\{\tilde{X}_{M},\tilde{X}_{N}\}=-(1+\frac{1}{2}P^{2})(X_{M}P_{N}-X_{N}P_{M})
\tag{3.14c}
\end{equation}
We see from the above equations that, on the constraint surface defined by
the first-class constraints (3.6)-(3.8), brackets (3.14) reduce to 
\begin{equation}
\{\tilde{P}_{M},\tilde{P}_{N}\}=0  \tag{3.15a}
\end{equation}
\begin{equation}
\{\tilde{X}_{M},\tilde{P}_{N}\}=\eta _{MN}-P_{M}P_{N}  \tag{3.15b}
\end{equation}
\begin{equation}
\{\tilde{X}_{M},\tilde{X}_{N}\}=-(X_{M}P_{N}-X_{N}P_{M})  \tag{3.15c}
\end{equation}
where, as in the massless particle case, the property (2.22) of the Dirac
bracket was used.

Now, keeping the same order of approximation used to arrive at brackets
(3.13), transformation equations (3.12a) and (3.12b) can be written as 
\begin{equation}
\tilde{X}^{M}-X^{M}=C_{i}^{M}(X,P)\phi _{i}  \tag{3.16a}
\end{equation}
\begin{equation}
\tilde{P}_{M}-P_{M}=D_{M}^{i}(X,P)\phi _{i}  \tag{3.16b}
\end{equation}
with $C_{1}^{M}=X^{M},$ $C_{2}^{M}=C_{3}^{M}=0$ and $D_{M}^{1}=-P_{M},$ $%
D_{M}^{2}=D_{M}^{3}=0$. Equations (3.16) are again in the form (2.14) and so
we can write 
\begin{equation}
\tilde{X}^{M}\approx X^{M}  \tag{3.17a}
\end{equation}
\begin{equation}
\tilde{P}_{M}\approx P_{M}  \tag{3.17b}
\end{equation}
Using these weak equalities in brackets (3.15) we rewrite them as

\begin{equation}
\{P_{M},P_{N}\}\approx 0  \tag{3.18a}
\end{equation}
\begin{equation}
\{X_{M},P_{N}\}\approx \eta _{MN}-P_{M}P_{N}  \tag{3.18b}
\end{equation}
\begin{equation}
\{X_{M},X_{N}\}\approx -(X_{M}P_{N}-X_{N}P_{M})  \tag{3.18c}
\end{equation}
Brackets (3.18) are the $(d+2)$-dimensional extensions of the $d$%
-dimensional brackets (2.27) we found for the massless particle. But now we
have a larger gauge invariance and so we can explicitly check the
observations we made at the end of the previous section about dual
gravitational fields.

We can now perform the gauge duality transformation 
\begin{equation}
X_{M}\rightarrow P_{M}  \tag{3.19a}
\end{equation}
\begin{equation}
P_{M}\rightarrow -X_{M}  \tag{3.19b}
\end{equation}
\begin{equation}
\lambda _{1}\rightarrow \lambda _{3}  \tag{3.19c}
\end{equation}
\begin{equation}
\lambda _{2}\rightarrow -\lambda _{2}  \tag{3.19d}
\end{equation}
\begin{equation}
\lambda _{3}\rightarrow \lambda _{1}  \tag{3.19e}
\end{equation}
under which, after an integration by parts, the Lagrangian in action (3.4b)
transforms as $\delta L=-\partial _{\tau }(X.P)$. Transformation (3.19)
therefore leaves action (3.4b) invariant. But transformation (3.19b) changes
the gauge function $\beta =\phi _{1}=\frac{1}{2}P^{2}\approx 0$ we used to
arrive at brackets (3.18) into the new gauge function $\beta =\phi _{3}=%
\frac{1}{2}X^{2}\approx 0$. Repeating the same steps as before for this
choice of $\beta $, and using again properties (2.14) and (2.22), we arrive
at the brackets 
\begin{equation}
\{P_{M},P_{N}\}=X_{M}P_{N}-X_{N}P_{M}  \tag{3.20a}
\end{equation}
\begin{equation}
\{X_{M},P_{N}\}=\eta _{MN}+X_{M}X_{N}  \tag{3.20b}
\end{equation}
\begin{equation}
\{X_{M},X_{N}\}=0  \tag{3.20c}
\end{equation}
We now clearly see that when the space-time coordinates have a vanishing
classical bracket, which will correspond to a commutative continuous
space-time in the quantized theory, a position-dependent effective
gravitational field given by 
\begin{equation}
G_{MN}=\eta _{MN}+X_{M}X_{N}  \tag{3.21}
\end{equation}
appears on the right side of equation (3.20b). This is the only possibility
for a physical gravitational field in a commutative space-time since,
according to equation (3.20a), the momenta will be noncommutative and
therefore discontinuous in the quantized theory.

We see from brackets (3.18) and (3.20) that the Minkowski metric tensor $%
\eta _{MN}$ plays the role of the unit tensor in both the space of the $%
G_{MN}(X)$ and the space of the $G_{MN}(P)$.

\section{ Concluding \ remarks}

In this work we proposed a canonical relation between gravity and space-time
noncommutativity. According to this canonical relation, in a noncommutative
space-time a physical gravitational field can not depend on the particle's
positions because this would imply sudden discontinuities in the field. In
this kind of space-time a physical gravitational field can depend only on
the particle's momenta, which are the continuous canonical variables. This
situation is inverted in a commutative space-time because now the momenta
are the discontinuous canonical variables.

\noindent

\noindent


\bigskip

\begin{thebibliography}{99}
\bibitem{1}  E. Witten, Nucl.Phys. B268 (1986) 253

\bibitem{2}  R. V. Mendes, J. Phys. A27 (1994) 8091

\bibitem{3}  C. Chryssomalakos and E. Okon, Int. J. Mod. Phys. D13 (2004)
2003 (hep-th/0410212)

\bibitem{4}  H. S. Snyder, Phys. Rev. 71 (1947) 38

\bibitem{5}  D. V. Ahluwalia, Phys. Lett. B339 (1994) 301 (hep-th/9308007)

\bibitem{6}  M. Buri\'{e}, T. Grammatikopoulos, J. Madore and G. Zoupanos,
JHEP 0604 (2006) 054 (hep-th/0603044)

\bibitem{7}  C. Leiva and M. Plyushchay, Ann. Phys. 307 (2003) 372
(hep-th/0301244)

\bibitem{8}  P. A. M. Dirac, \textsl{Lectures on Quanum Mechanics}, Yeshiva
University, 1964

\bibitem{9}  M. Henneaux and C. Teitelboim, \textsl{Quantization of Gauge
Systems}, Princeton University Press 1992

\bibitem{10}  I. Bars and C. Kounnas, Phys. Lett. B402 (1997) 25
(hep-th/9703060)

\bibitem{11}  I. Bars, C. Deliduman and O. Andreev, Phys. Rev. D58 (1998)
066004 (hep-th/9803188)

\bibitem{12}  I. Bars and C. Kounnas, Phys. Rev. D56 (1997) 3664
(hep-th/9705205)

\bibitem{13}  I. Bars, hep-th/9809034

\bibitem{14}  I.Bars, C. Deliduman and D. Minic, Phys. Lett. B 466 (1999)
135 (hep-th/9906223)

\bibitem{15}  I. Bars, Phys. Rev. D58 (1998) 066006 (hep-th/9804028)

\bibitem{16}  I. Bars, Phys. Rev. D62 (2000) 085015 (hep-th/0002140)
\end{thebibliography}
\end{document}